\documentclass[12pt]{article}
\usepackage{epsfig,graphpap,times}
\usepackage{amsmath}
\usepackage{amssymb}
\usepackage{amsthm,txfonts}
\usepackage[round,numbers,sort&compress]{natbib}
\usepackage{color}
\def\rbuildrel#1\over#2{\mathrel{\mathop{#2}\limits_{#1}}}

\catcode`@=11
\def\underline#1{\relax\ifmmode\@@underline#1\else
        $\@@underline{\hbox{#1}}$\relax\fi}

\def\titlepage{\pagestyle{empty}\c@page=0
      \def\thefootnote{\fnsymbol{footnote}} }
\def\endtitlepage{\pagestyle{plain}\c@page=1
      \def\thefootnote{\arabic{footnote}} \c@footnote\z@ }
\catcode`@=12

\newskip\humongous \humongous=0pt plus 1000pt minus 1000pt

\newif\ifdtup

\def\*{\hskip .06 cm}
\def\thebibliography#1{\section*{\ \markboth
 {REFERENCES}{REFERENCES}}\list
 {[\arabic{enumi}]}{\settowidth\labelwidth{[#1]}\leftmargin\labelwidth
 \advance\leftmargin\labelsep
 \usecounter{enumi}}
 \def\newblock{\hskip .11em plus .33em minus -.07em}
 \sloppy
 \sfcode`\.=1000\relax}

\newcommand{\beq}{\begin{equation}}
\newcommand{\enq}{\end{equation}}
\newcommand{\bqa}{\begin{eqnarray}}
\newcommand{\eqa}{\end{eqnarray}}
\newcommand{\bqas}{\begin{eqnarray*}}
\newcommand{\eqas}{\end{eqnarray*}}
\newcommand{\bea}{\begin{array}}
\newcommand{\ena}{\end{array}}

\date{}
\begin{document}

\begin{center}
\noindent{\large \bf Calculating potential of mean force between like-charged nanoparticles: a comprehensive study on salt effects}\\
\end{center}
\vskip .1 in
\begin{center}
{\bf Yuan-Yan Wu$^{\dag}\,\,$    Feng-Hua Wang$^{\dag}\,\,$    and 
 Zhi-Jie Tan$^{\star}$}\\

Department of Physics and Key Laboratory of Artificial Micro \& Nano-structures of Ministry of Education,
School of Physics and Technology, Wuhan University, Wuhan, P.R. China 430072\\

\end{center}

\vskip .2 in
\centerline{\large \bf Abstract}
\vskip .1 in

\normalsize{ Ions are critical to the structure and stability of polyelectrolytes such as nucleic acids. In this work, we systematically calculated the potentials of mean force between two like-charged nanoparticles in salt solutions by Monte Carlo simulations. The pseudo-spring method is employed to calculate the potential of mean force and compared systematically with the inversed-Boltzmann method. An effective attraction is predicted between two like-charged nanoparticles in divalent/trivalent salt solution and such attraction becomes weakened at very high salt concentration. Our analysis reveals that for the system, the configuration of ion-bridging nanoparticles is responsible for the attraction, and the invasion of anions into the inter-nanoparticles region at high salt concentration would induce attraction weakening  rather than the charge inversion effect. The present method would be useful for calculating effective interactions during nucleic acid folding.\\

}

\noindent -----------------------------------------------------------------------------------------------------------\\
\noindent{$^{\dag}$The authors contributed equally to the work.}\\

\noindent{$^{\star}$Corresponding author: zjtan@whu.edu.cn}\\


\vfill\eject
%\endtitlepage
\topmargin = -.2 in

\section{Introduction}

Ions play critical roles in the thermodynamic and kinetic properties of charged systems, such as nucleic acids and other polyelectrolytes \cite{1,2,3,4,5,woodson,6,7}. For example, the folding of nucleic acids into compact native structures would bring the build up of negative charges and  requires cations to neutralize the negative backbone charges, and for other polyelectrolytes, ionic condition is also essential for their structure and stability \cite{8,9,10,12,13,14,15,AJLiu2Macro,AJLiu1Pre}.\\
\indent Extensive experiments, theories and simulations have been employed, attempting to obtain a fundamental understanding on the important and complex roles of ions in polyelectrolyte system \cite{PNAS2010Ritort,16,17,18,19,PRE2010xuzhenli,20,Jungwirth2008,langmuir2012Martin}. A typical paradigm is the system of two like-charged polyelectrolytes immersed in ion solution, which has attracted considerable interests in recent years, since ions can switch the like-charge repulsion into attraction at some ionic conditions \cite{21,22,23,JCP2006Jonsson}. For DNA and RNA, the addition of multivalent salt ions can condense DNA/RNA from extended state to compact state \cite{24,25,26,27,28,PNAS2005Wong,HLowen1,HLowen2Pre}. For (spherical) polyelectrolytes, the multivalent ions could induce the dispersed distribution to form compact clusters \cite{29,30,31,32,LPollack}. Although many efforts have been made and much progress has been achieved on understanding the ion roles in modulating like-charge interaction \cite{
PRE2001Linse,JCP1984Jonsson,JPCM2002Linse}, there is still lacking of the comprehensive understanding on how ions influence the effective interactions, including the effects of ion concentration, valence, size and charge density, especially at very high salt and charge density on polyelectrolytes.\\
\indent However, to quantify the ion effects in the systems of highly charged polyelectrolytes still remains a challenge, especially for multivalent ions, not only because the ion-modulated interaction is not strong ($\sim$ $ k_{B}T $), but also polyelectrolytes could induce (strong) correlations between (multivalent) ions in the vicinity of molecular surface. Up to now, there have been two classic theories for treating ion-polyelectrolyte interaction: the counterion condensation (CC) theory \cite{33} and the Poisson-Boltzmann (PB) theory \cite{34,35,36,37,38,39,40,41}. The two theories are rather successful in predicting electrostatic properties of polyelectrolyte in monovalent/aqueous solutions. Nevertheless, the CC theory is based on the line-charge structural model and is developed for dilute salt solution and linear polyelectrolytes of infinite length. Thus it is inapplicable for shaped polyelectrolytes in salt solutions. The PB theory is based on the Poisson equation with a Boltzmann weighted mean
distribution for diffusive ions, where ions are modeled as continuous fluid-like particles moving independently in a mean electrostatic field. Thus, the PB theory ignores discrete ion properties such as ion correlation and ion fluctuation, and  always predicts like-charge repulsion even in multivalent ion solutions. For spherical polyelectrolytes, the Derjaguin-Landau-Verwey-Overbeek (DLVO) theory has been developed to describe the effective interactions \cite{42,43,44}. The electrostatic contribution of the DLVO interactions comes from the linearization of nonlinear term in PB equation at weak electric potential approximation, and has the form of Debye-H\"{u}ckel-type potential for two spherical polyelectrolytes. In analogy to PB, the DLVO theory always predicts a (screened) like-charge repulsion.\\
%%\indent In the recent years, various improvements on the CC and the PB theories are being performed continuously  \cite{EPJE2011manning,PRE2003Shiessel,PRE2009HCChang}.
\indent To account for the effects of ion correlation and ion-binding fluctuation, some advanced theories have been proposed, such as the dressed ion theory/strong coupling theory for charged colloids \cite{45,46,47}, integral equation theory for polyelectrolyte \cite{48,50}, and tightly bound ion theory for nucleic acids \cite{51,52,53,54}. These advanced theories have successfully predicted a variety of thermodynamic properties for polyelectrolytes and nucleic acids in ionic solutions, while they are either developed for the salt-free solutions in strong coupling limit or for specified polyelectrolytes (e.g., nucleic acids). To deal with the electrostatic properties of  biomolecule system, some other approaches have been developed,  such as variational multiscale models and density-functional theory \cite{guoweiWei,JZWuP}. As a needful bridge between theories and experiments, computer simulations have become a powerful tool for the multi-body statistical systems and has 
made many valuable predictions for charged systems \cite{Holm2005,55,56,57}. Especially in recent years, along with the great development of computation facility, all-atom MD simulations have been used to predict the interaction between biological molecules. Recently, Luan \textit{et al} predicted the interaction between two short DNA helices in monovalent and divalent electrolytes by the all-atom MD simulation, as well as the end-to-end interaction between DNAs and DNA-DNA interaction in tight supercoils \cite{luan JACS,luan PRL,luan Soft Matter,luan NAR}. The \textquotedblleft ion bridge"  has been proposed to be responsible  for the effective attraction between two like charge DNAs.\\
\indent In this paper, we will employ Monte Carlo (MC) simulations to systematically calculate the ion-modulated potential of mean force (PMF) between two like-charged nanoparticles. The pseudo-spring method is employed to calculate the PMF, and compared systematically with the inversed-Boltzmann method. Beyond the previous studies, \textit{we emphasize the method for calculating the PMF, and physical mechanism for the ion modulated like-charge interactions at extensive ionic conditions, especially at very high salt concentration.} Such methods would be helpful for probing the effective interaction during nucleic acid folding.

\section{Model and methods}

In this work, for simplicity, we use charged macro-spheres to represent nanoparticles \cite{PRE2006holm,PRE2006Wong}, and the ion solution is considered to be an ensemble of small spheres of different charges and sizes, different kinds of ions are represented  by the corresponding  charge and size, and all of them are dispersed in a continuum dielectric medium whose permittivity corresponds to that of the solvent.\\
\indent The interaction defining the system is composed of two contributions: the electrostatic interaction $ U_{\text{el}} $, and the excluded volume interaction $ U_{\text{ex}} $. The electrostatic interaction $ U_{\text{el}} $ between charges $ i $ and $ j $ (ions and nanoparticles) is given by
\begin{equation}\label{Eq.2}
  \mathrm{\emph{ U}_{el}=\frac{\emph{q}_{i}\emph{q}_{j}}{4\pi\epsilon\epsilon_{0}\emph{r}}} \,\, ,
\end{equation}
where $ q_{i} $ and $ q_{j} $ are charges on spheres $ i $ and $ j $, and $ r $ is the center-to-center distance between the two spheres. $ \epsilon $ is the dielectric constant of solvent ($ \epsilon=78 $ at room temperature), and $ \epsilon_{0} $ is the permittivity of vacuum. The excluded volume interaction $ U_{\text{ex}} $ between spheres $ i $ and $ j $ is accounted for by a repulsive Lennard-Jones potential
\begin{equation}\label{Eq.3}
   \mathrm{ \textit{U}_{ex}=\begin{cases}{}
           4 U_{0}((\frac{\sigma}{r})^{12}-(\frac{\sigma}{r})^{6})\qquad \textrm{for} \,\, r<\sigma \, ;  \\
           0 \qquad \qquad \qquad \qquad \,\,\,\,\,\,\,\,\, \textrm{for} \,\, r\geq\sigma \, ,
             \end{cases}}
\end{equation}
where $ \sigma $ is the sum of the radii of the two spheres, and $ U_{0} $ is the volume exclusion strength. In this study, we take $ U_{0}=100 $. Our control test shows that our results are not sensitive to the value of $ U_{0} $ around 100. \\
\indent The simulation cell is a rectangular cell where periodic boundary condition is applied. To diminish the boundary effect, we always keep the cell size larger than two nanoparticles by six times of the Debye length, and the calculation results are stable as tested against different cell sizes. In the simulations, the radii of nanoparticles and ions are taken as 10{\AA} and 2{\AA}, respectively. The charges $Z$ on nanoparticles are taken as $-21e$, ensuring the surface charge density is close to that of phosphate groups in nucleic acid \cite{11}.  Also, the additional calculations are performed for other ion radii (3{\AA}, 4{\AA}) to study the ion size effect, and for other nanoparticle charge $Z$ $(=-12e, -24e, -36e)$ to study the charge density effect. In the calculations of potential of mean force $ \Delta G(x)=G(x)-G(x_{\textrm{ref}}) $, for simplicity, the outer reference distance $ x_{\textrm{ref}} $ is taken as 40{\AA} for all ion conditions.\\
\indent We used the Metropolis Monte Carlo algorithm for all simulations in our work, which is a computational approach for generating a set of $N$ configurations of the system by the relative probability proportional to the Boltzmann factor: $p(N_i)\varpropto e^{-E(N_i)/k_B T}$, and the transition probability $p_{i\rightarrow j}$ from configuration $ i $ to configuration $  j $ is given by $p_{i\rightarrow j}=e^{-(E_j-E_i)/k_B T}$. Starting from an initial configuration with the two nanoparticles in the center and the ions randomly distributed in the simulation box, every particle randomly moves to a trial position and we calculate the energy change $\Delta E$ due to the move. If a random number $R(\in[0,1])<p=e^{-\Delta E/k_B T} $, the trial move is accepted. Repeat the trial move until the system reaches the equilibrium. Fig. S1 and Fig. S2 (in the Supporting Material) shows that the statistical results of our simulations can quickly reach the convergence.\\
\indent Based on the MC simulations with the above described system and energy functions, we calculate the potential of mean force for two nanoparticles with the pseudo-spring method, as well as the inversed-Boltzmann method.\\

\subsection{Pseudo-spring method}

To calculate the PMF between two nanoparticles, we add a pseudo-spring with spring constant $ k $ to link the centers of the two nanoparticles as shown in Fig. 1a. The effective force between the two nanoparticles can be given by
\begin{equation}\label{Eq.4}
F=k\Delta x ,
\end{equation}
 where $ \Delta x $ is the deviation of the spring length away from the original length $ x_{0} $ at equilibrium. Then the PMF between the two nanoparticles can be calculated by the integration
\begin{equation}\label{Eq.5}
 \Delta G(x)=G(x)-G(x_{\textrm{ref}})=\int^{x}_{x_{\textrm{ref}}}F(x')dx' \, .
\end{equation}\\
  Eq. 4 shows that we need to calculate $F(x)$'s at a series of $x$ in order to obtain $\Delta G(x)$.\\
 \indent Figs. 1a-c illustrate the process of calculating PMF between the two nanoparticles using the pseudo-spring method. Firstly, we employed MC simulation for the system of pseudo-spring linked nanoparticles in salt solution. The statistical analysis on the fluctuation of $ x $ \textit{versus} MC steps at equilibrium (shown Fig. 1b) gives the distribution probability $p(x) $ of separation $ x $, which can be used to estimate $ \Delta x $. In practice, we fit the distribution probability to the Gaussian function \( g(x)\propto e^{-\frac{(x-b)^{2}}{2c^{2}}} \) to obtain the deviation in spring length $ \Delta x(=b-x_{0}) $. The negative and positive $\Delta x$'s correspond to the attractive and repulsive effective force, respectively. Consequently, the force $ F(x) $ and PMF $ \Delta G(x) $ between the two nanoparticles can be  calculated based on $ \Delta x $'s and the above described formulas (Eq. 3 and 4).\\

 \subsection{ Inversed-Boltzmann method}

 As a comparison for the above pseudo-spring method, we also briefly introduce the inversed-Boltzmann method here. For the system of two nanoparticles in salt solution, we use MC simulation to get the ensemble of configurations of the two nanoparticles and ions. The statistics on $ x $ at equilibrium give the distribution probability $ p(x) $ of separation $ x $ between two nanoparticles. Since the distribution probability $ p(x) $ at equilibrium satisfies the Boltzmann distribution $ p(x)\sim e^{-G(x)/k_{B}T} $, the potential of mean force can be calculated by
\begin{equation}\label{Eq.6}
 \Delta G(x)=G(x)-G(x_{\textrm{ref}})=-k_{B}T\ln\frac{p(x)}{p(x_{\textrm{ref}})} \, .
\end{equation}
As an example, Fig. 1d-f shows how the method is employed to calculate the PMF between two nanoparticles.

\section{Results and discussion}

 In the work, we calculated the PMF \(\Delta G(x)\) between nanoparticles by the pseudo-spring method,  as well as the inversed-Boltzmann method, and examined how the extensive ionic conditions modulate the PMF. \textit{We emphasize the methods and the mechanism at various ionic conditions, especially at high salt concentration.} For convenience, we use [1+], [2+] and [3+] to stand for the 1:1, 2:2 and 3:3 salt concentrations, respectively.

\subsection{Like-charge repulsion to attraction modulated by ion valence}

\noindent{\emph{ \text{Potential of mean force modulated by ion valence.}}} \\
\indent  Figs. 2a-c show that monovalent and multivalent ions have the contrasting effect on the PMF between two like-charged nanoparticles. For monovalent ions, the PMF is repulsive. As ion valence increases, the repulsive PMF between nanoparticles transits to the attractive one, and the effective attraction becomes stronger for higher valence. For example, at 1mM 2:2 salt, the minimum of PMF has the value of  $\sim$ $-0.8k_{B}T$, while at 1mM 3:3 salt, the minimum of PMF is $\sim$ $-3.8 k_{B}T$.
Such transition from repulsive PMF to attractive one is coupled to the ion valence. Higher valent ions can interact more strongly with nanoparticles and the entropic penalty for ion-binding is much lower, which is responsible for the attractive PMF in multivalent salt solution and will be discussed in details in the following subsection. \\
\indent  Figs. 2a-c also show that the pseudo-spring method gives nearly identical predictions on the PMF to those from the inversed-Boltzmann method, which suggests that our predictions are robust. It is also noted that the pseudo-spring method makes reliable prediction for all separations $ x $ over the wide ion concentration range, while the inversed-Boltzmann method can not give prediction for small $ x $ at low 1:1 salt in a
reasonable computation time. This is attributed to the very high free energy barrier for small $x$ at low 1:1 salt which corresponds to the very low probability that can be rarely sampled in the simulations without spring constraint. \\
\indent The predictions from the DLVO theory are also shown in the same figures as a comparison. For high 1:1 salt, the DLVO prediction agrees well with that from the MC simulations, while for low 1:1 salt, the DLVO overestimates the repulsive interactions. For 2:2 and 3:3 salts, the DLVO theory predicts the repulsive interaction, which is qualitatively different from the simulations. Such discrepancy comes from the two approximations made in the DLVO: the weak-field approximation and the mean-field approximation. For the strong electrostatic field, the nonlinear term in PB equation cannot linearized \cite{58}, and thus DLVO theory overestimates the electrostatic repulsion for low 1:1 salt. Furthermore, the neglect of ion correlations in PB (and DLVO) theory would neglect the organization of binding ions between two nanoparticles which can form the ion-configurations favorable for an effective attraction, and thus DLVO only predicts a reduced like-charge repulsion.\\

\noindent{\emph{\text{Driving force for the effective like-charge attraction.}} }\\
\indent To gain a deep understanding about the ion effect and the driving force for the ion-mediated like-charge attraction, the PMF \(\Delta G\) is decoupled into two components: the electrostatic energy \(\Delta G_{E}\) and the entropic free energy \(\Delta G_{S}\) by \cite{59}
\begin{equation}\label{Eq.7}
 \Delta G_{E}(x)=\langle U_{E}(x)-U_{E}(x_{\textrm{ref}}) \rangle_{\textrm{conf}} ;
\end{equation}
\begin{equation}\label{Eq.8}
 \Delta G_{S}(x)=\Delta G(x)-\Delta G_{E}(x) ,
\end{equation}
where \( U_{E}(x) \) is the total Coulomb energy at separation \( x \), and  \( \langle  \rangle_{\textrm{conf}} \) denotes the averaging over all the ion distribution configurations at equilibrium.\\
\indent Figs. 3a-c show that, as the two nanoparticles approach each other from $x_{\textrm{ref}}$ (40{\AA}), \(\Delta G_{E}\) decreases, while \(\Delta G_{S}\) increases monotonously. Thus, \(\Delta G_{E}\) and \(\Delta G_{S}\) tend to give an attractive and repulsive forces, respectively. Therefore, the driving force for the effective attraction comes from the electrostatic energy \(\Delta G_{E}\). Physically, the decrease (increase) of \(\Delta G_{E}\) (\(\Delta G_{S}\)) with the decrease of $ x $ is attributed to the ion binding to the nanoparticles. Here, we use $Q(r)$ to represent the net ion charge fraction within a distance $r$ from the nanoparticles.
 \begin{equation}\label{Eq.9}
  Q(r)=\int_{<r}\sum _{\alpha}Z_{\alpha}d^{3}\textbf{r}
\end{equation}
where $Z_{\alpha}$ denotes the valence of $\alpha$ ion species. As shown in Figs. 2d-f, when $x$ is decreased from 40{\AA} to 25{\AA}, more cations bind to the nanoparticles, causing a stronger charge neutralization. The more binding ions interacting with nanoparticles and the correlations between them cause the decrease of \(\Delta G_{E}\)  with the decreased \(x\). Simultaneously, more binding ions bring the higher ion-binding entropic penalty, causing the increase of \(\Delta G_{S}\). The competition between the decreasing (attractive) \(\Delta G_{E}\) and the increasing (repulsive) \(\Delta G_{S}\) results in the overall \(\Delta G(x)\).\\
\indent To analyze the effect of ion valence on \(\Delta G_{S}\) and \(\Delta G_{E}\), we use \(\Delta Q\) to represent the increase of binding ions around the nanoparticles as they approach each other. As shown in Figs. 2d-f, \(\Delta  Q\) for low valent salt is larger than that for high valent salt: $\Delta Q_{1+}>\Delta Q_{2+}>\Delta Q_{3+}>0$, and correspondingly, the increase in \(\Delta G_{S}\) is much stronger for low valent salt: $\Delta G_{s}(1+)>\Delta G_{s}(2+)>\Delta G_{s}(3+)$. As the result, the strong (weak) repulsive \(\Delta G_{S}\) and attractive \(\Delta G_{E}\) give the distinct overall repulsive/attractive effective interaction for monovalent/multivalent salts, respectively.\\
\indent To directly show how binding ions correlate with the nanoparticles and induce the negative \(\Delta G_{E}\), we plot the ion charge density and the snapshots for the nanoparticles with binding ions at two separations \(x=40\){\AA} and \(25\){\textrm{\AA}}, respectively. As shown in Fig. 4, as \( x \) decreases, ions like to reside in the region between the nanoparticles and these ions are shared by both of the two nanoparticles. Such ion-bridging-nanoparticle configuration causes the decrease of \(\Delta G_{E}\) with \( x \). Although \(\Delta Q\) is less strong for multivalent ions, the organization of ion coordinates in the inter-nanoparticle region can still cause the strong decrease in \(\Delta G_{E}\).

\subsection{Non-monotonous behavior of PMF \textit{versus} ion concentration }

\noindent{\emph{\text{Repulsion in 1:1 salt weakened by higher ion concentration. }}} \\
\indent As shown in Fig. 2a, for 1:1 salt, \(\Delta G(x)\) increases monotonously for the two approaching nanoparticles no matter what [1+] is. As [1+] is increased, the repulsive PMF becomes weaker. This is attributed to the decreased entropy penalty for ion binding and the stronger ion neutralization at higher [1+]; see Fig. 2d for the ion-binding numbers \(\Delta  Q\) and Fig. 3a for \(\Delta G_{S}\).\\

\noindent{\emph{\text{Attraction in 2:2 and 3:3 salts enhanced by higher ion concentration.  }}}\\
\indent For 2:2 salt, as shown in Fig. 2b, at low [2+] ($\sim$0.1mM), the effective interaction between the two nanoparticles is weak until \(x\leq24{\textrm{\AA}}\). As [2+] is increased, the nanoparticles tend to attract each other, and such attraction becomes stronger. Similarly, the attractive PMF also becomes stronger for higher [3+].\\
\indent The above described [2+]/[3+]-dependence is attributed to the decreased ion-binding penalty and increased ion-binding number at higher [2+]/[3+]. As shown in Figs. 2ef, for higher [2+]/[3+], more ions strongly bind to the nanoparticles even at \(x=x_{ref}\), and the increase in binding ion number, \(\Delta Q\), due to nanoparticles approaching is weaker, causing lower \(\Delta G_{S}\). At the same time, owing to the stronger ion correlations accompanying with the nanoparticles approaching, \(\Delta G_{E}\) decreases. The attractive (negative) \(\Delta G_{E}\)  and the much weaker repulsive (positive) \(\Delta G_{S}\) give the overall stronger attractive PMF at higher [2+]/[3+], as shown in Figs. 3bc.\\

\noindent{\emph{\text{Attraction weakening at very high 2:2 and 3:3 salts.}} }\\
\indent However, as [2+] or [3+] continues to increase and exceeds a certain critical value \(c^{\ast}\), the further addition of salt would weaken the effective attraction. For example, as shown in Fig. 5a, when [2+] is increased from 0.001M to 0.1M, the attraction becomes stronger gradually. However, when [2+] is increased to 0.125M/0.3M, the effective attraction becomes apparently weaker than that at 0.1M. Also as shown in Fig. 5d,  the effective attraction is also apparently weakened when [3+] exceeds $\sim$1mM. Such phenomena of attraction weakening may correspond to the resolubility of polyelectrolyte aggregates such as DNA and F-actin \cite{60,61}. What is responsible for the attraction weakening at high salt? Is it caused by the over-neutralization by binding ions?\\
\indent To answer the question, we examine the ion-binding/distribution near the nanoparticles. As shown in Figs. 5be, for higher salt, more ions would bind to the nanoparticles, resulting in possible full-neutralization or over-neutralization \cite{prl2000ayGrosberg,jcp2001ayGrosberg,wfh}. For divalent ions, from 0.001M to 0.1M, the binding ions increase gradually, and reach near full-neutralization (\(Q\sim1\)) at \(\sim\)0.1M, and there is almost no apparent change on $Q(r)$ from 0.1M to 0.125M. At \(\sim\)0.3M, the weak and visible over-neutralization appears in the region closely around the nanoparticles. Similarly, for trivalent ions, from 0.01mM to 1mM, the number of binding ions increases gradually, and the over-neutralization appears when [3+] exceeds \(\sim\)10mM. Figs. 5abde, show that, the (apparent) over-neutralization is not definitely required to achieve the weakening of the effective like-charge attraction. Because the attraction, is notably weakened while there is no visible over-
neutralization, when [2+] is increased from 0.1M to 0.125M.\\
\indent To get a deep understanding for such attraction weakening, we focus on the ions between the two nanoparticles (the shaded region shown in Fig. 6a) rather than the global binding ions. What happened in this correlation region when the attraction becomes weakened? As shown in Fig. 5c, for 2:2 salt, as [2+] is increased from 0.0001M to 0.3M, the net ion charge $Q_{T}$ in the inter-nanoparticles region only has a very slight increase. Simultaneously, the anion charge $Q_{2-}$ increases very weakly from 0.0001M to 0.01M (with value of \(\sim\)0), while increases sharply when [2+] exceeds $\sim$0.1M. The same phenomena also appears for 3:3 salt. As shown in Fig. 5f, the net ion charge $Q_{T}$ changes very weakly over the wide [3+] covered here, while the anion charge $Q_{2-}$ increases sharply from \(\sim\)0 when [3+] exceeds $\sim$3mM. The transition points of \(\sim\)0.1M for divalent ions, and \(\sim\)3mM for trivalent ions just correspond to the transition concentrations $c^{\ast}$ for the attraction
weakening for divalent and trivalent ions
respectively. \textit{Such sharp increase of the number of anions between nanoparticles from \(\sim\)0 at the transition points of the attraction weakening may suggest that the invasion of anion into the inter-nanoparticle region is responsible for the attraction weakening at high 2:2/3:3 salts.}\\
\indent To confirm the supposition, we make the additional test simulations for divalent and trivalent solutions respectively, by restricting a anion in the inter-nanoparticle region, see Fig. 6a. As shown in Fig. 6b for 0.1M [2+] salt and Fig. 6c for 1mM [3+] salt, the restriction of an anion between the nanoparticles causes the visible attraction weakening for both 2:2 and 3:3 salts. Based on the above phenomena and the test calculations, we conclude that, the invasion of anions into the inter-nanoparticles region is responsible for the attraction weakening at high multivalent salts rather than the global charge over-neutralization effect.

\subsection{Ion size effect }

Several experiments have demonstrated the important role of ion size in the compaction of polyelectrolytes such as RNA/DNA, rodlike M13 and fd viruses \cite{62,63,64,65}. In this section, we focus on the physical mechanism for divalent ion size effect in the PMF. Specifically, we use different ion radii of 2.0{\AA}, 3.0{\AA}, and 4.0{\AA}. As shown in Fig. 7a, when ion size is decreased from \(4{\textrm{\AA}}\) to $2{\textrm{\AA}}$, the predicted PMF changes from a (weakly) repulsive one to an attractive one with a smaller separation and a lower minimum PMF (\(\Delta G_{\textrm{min}}\)), suggesting a stronger attraction for smaller ions.\\
\indent Smaller ions can make closer contact and stronger interaction with the nanoparticles. Such stronger interaction would cause a stronger charge neutralization for the nanoparticles, as shown in Fig. 7b. As the nanoparticles approach each other, the increase in the number \((\Delta Q)\) of binding ions and the resultant \(\Delta G_{S}\) are (slightly) smaller for small ions than for larger ions (see Fig. 7c). Simultaneously, the more binding ions interacting with nanoparticles still results in the negative \(\Delta G_{E}\) and \(\Delta G_{E}\) decreases with \(x\) to a smaller \(x\) for small ions, due to stronger Coulomb interaction with nanoparticles and the smaller ion  volume exclusion. At very small \(x\), the binding ions can be pushed out from the strongly correlated inter-nanoparticles region and \(\Delta G_{E}\) would increase. Such effect is stronger for more bulky ions because of the larger ion excluded volume. As the result, the nanoparticles are more strongly/closely stabilized by smaller
ions.

\subsection{Effect of charge density of nanoparticles }

To examine the effect of nanoparticle charge density, we use different charges \(Z\) on the nanoparticles: \(Z=-12e, -24e\) and \(-36e\). Fig. 8a shows the PMF per unit charge \(\Delta G(x)/Z\) for different $Z$'s in 0.01M 2:2 salt solution. For low charge density \( (Z=-12e) \), the PMF between the nanoparticles is repulsive. With the increase of $Z$ to -36e, the PMF becomes attractive. Thus, higher \(Z\) enhances the effective attraction. To analyze the mechanism for such effect, we decouple the \(\Delta G(x)\) into \(\Delta G_{E}\) and \(\Delta G_{S}\) according to Eqs. 6 and 7.\\
\indent As shown in Fig. 8b, for higher $Z$, more ions become binding even at large $x$. As $x$ is decreased from 40{\AA} to 25{\AA}, the increase in binding ions (\(\Delta Q\)) follows the following order: $\Delta Q_{-12e}>\Delta Q_{-24e}>\Delta Q_{-36e}$. The electrostatic free energy per unit charge ($\Delta G_{E}/Z$) decreases rapidly as the two nanoparticles approaching, following the order of $\Delta G_{E}/Z(-12e)<\Delta G_{E}/Z(-24e)<\Delta G_{E}/Z(-36e)<0$, as shown in Fig. 8c. Such order is attributed to two reasons: $(\romannumeral1)$ there are apparently more ions binding to the nanoparticles of lower $Z$ as they approach each other, then the decrease of $\Delta G_{E}/Z$ for the nanoparticles of lower $Z$ is stronger than that for higher $Z$; $(\romannumeral2)$ as the nanoparticles become close, the more binding ions can correlate with the nanoparticles, to form low-energy states, resulting the decreased $\Delta G_{E}$. When the
nanoparticles become very close, $\Delta G_{E}$ increases due to the ion-nanoparticle volume exclusion. Simultaneously, corresponding to the much larger \(\Delta Q\) due to the two nanoparticles approaching, $\Delta G_{S}$ for the nanoparticles with low $Z$ is much stronger than that for those with high $Z$: $\Delta G_{S}/Z(-12e)>\Delta G_{S}/Z(-24e)>\Delta G_{S}/Z(-36e)>0$. The competition between $\Delta G_{E}$ and $\Delta G_{S}$, gives the overall PMF. Higher the charge density on nanoparticles is, stronger the effective attraction becomes due to the apparently lower entropic free energy $\Delta G_{S}$. For very low charge density, the PMF can become repulsive.

\subsection{Pseudo-spring method \textit{versus} inversed-Boltzmann method }

The two methods including pseudo-spring and inversed-Boltzmann methods were employed to calculate the PMFs between two like-charged nanoparticles. The comparisons between the two methods lead to the following major conclusions on the methods.
\begin{enumerate}
  \item On the prediction. The two methods nearly make the identical predictions on the PMFs except for low 1:1 salt.
  \item On the applicability. The pseudo-spring method works well over the wide ionic conditions, while the inversed-Boltzmann method only make reliable prediction for relatively ``flat'' PMF profiles rather than those with high free energy barrier.
  \item On the computation efficiency. Generally, for ``flat'' PMF, the pseudo-spring method is more efficient than the inversed-Boltzmann method by over 2-fold and the computation efficiency is not sensitive to the ``flat'' or ``steep'' PMF profile. For the case of low 1:1 salt, the inversed-Boltzmann method is very inefficient and possibly could not give reliable prediction in a reasonable time because of the high free energy barrier.
  \item On thermodynamic analysis. In pseudo-spring method, the two nanoparticles are constrained and only slightly fluctuated from the original coordinate, thus the method is practically convenient in the analysis on ion-binding and thermodynamics such as entropy and enthalpy.
\end{enumerate}
Overall, the pseudo-spring method is more reliable in the applicability, computation efficiency, and data analysis. Additionally, we have made a comparison between the pseudo-spring method and the weighted histogram analysis method  (WHAM) extensively used in all-atom MD simulation, and the two methods give the similar prediction (in the Supporting Material).

\section{ Conclusions and Discussion}

In this work, we employed the pseudo-spring method calculate the potential of mean force between two like-charged nanoparticles in monovalent, divalent and trivalent salt solutions, and analyzed the physical mechanism for the effective interaction between the two nanoparticles especially at very high salt concentration.  The following is a brief summary of the major findings.
\begin{enumerate} \setlength{\itemsep}{-0.5ex}
\item As the two like-charged nanoparticles approach to each other, the entropic free energy increases and tends to give a repulsive force, due to increased entropy penalty for more binding ions, while the electrostatic energy decreases and tends to give a attractive force due to more binding ions correlating with the two nanoparticles. The relative strengths of the two components  can be modulated by ion valence, concentration and size, and charge density on nanoparticles, thus give the overall (repulsive or attractive) PMF between two like-charged nanoparticles.
\item At very high 2:2 ($\sim$0.1M) and 3:3 ($\sim$3mM) salt concentrations, the attractive PMF can be weakened by the further addition of salt ions. Such attraction weakening at high salt concentration is attributed to the invasion of anions into the inter-nanoparticle region.
\item In calculating the effective interaction, the pseudo-spring method is more reliable in prediction efficiency, applicability and data analysis, as compared with the inversed-Boltzmann method.
\end{enumerate}
\indent $ \,\,\,\,\, $ The present model also involves some approximations and simplifications. First, the solvent (water) molecules are modeled implicitly as a uniform medium with dielectric constant of water, and thus, the entropy effect of water molecules is implicitly accounted for in electrostatic energy rather than in entropic free energy. Second, the dielectric discontinuity at the boundary between solvent and nanoparticles is ignored, which can affect the ion-binding in the vicinity of particle surface and needs to be taken into account in the future work. Finally, to simplify the computation complexity, we choose the $x=40{\textrm{\AA}}$ as the outer-reference separation to calculate the potential of mean force, and such simplification might slightly affect the results. Nevertheless, our model gives an overall picture for ion-modulated like-charge interaction and the method employed in the study can be useful for probing effective interaction during the folding of  biomolecules such as nucleic acids 
and proteins.

\section{ Acknowledgments}

We are grateful to Prof. Shi-Jie Chen for valuable discussions. This work was supported by the National Science Foundation of China grants (11074191 and 11175132), the Program for New Century Excellent Talents (NCET 08-0408), the Fundamental Research Funds for the Central Universities (1103007), the National Key Scientific Program (973)-Nanoscience and Nanotechnology (No. 2011CB933600) and by SPF for ROCS, SEM. One of us (Y.Y.Wu) also thanks financial supports from the interdisciplinary and postgraduate programs under the Fundamental Research Funds for the Central Universities.

\newpage

\begin{figure}[ht]
\begin{center}
\epsfxsize=18.0cm \centerline{\epsfbox{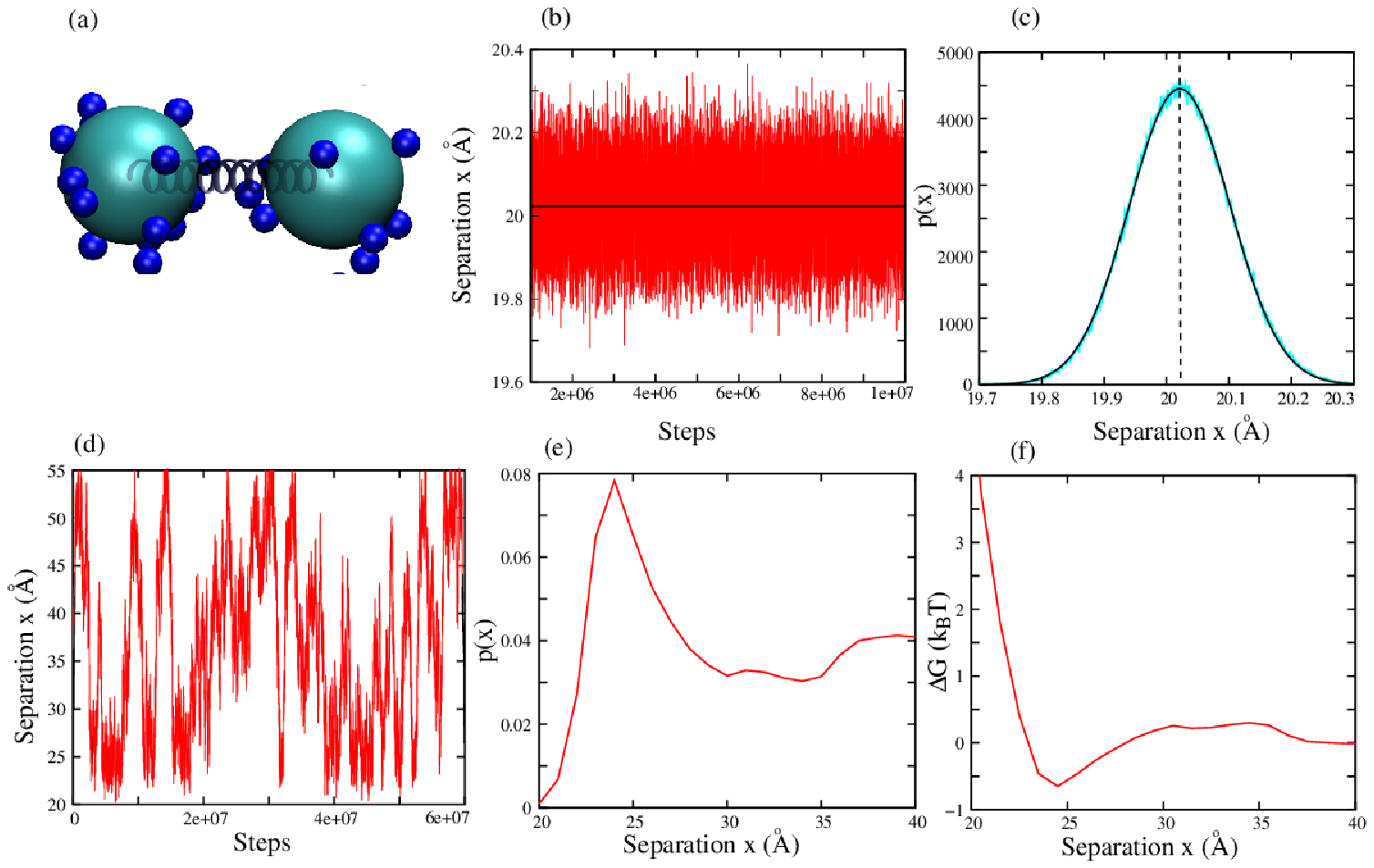}}
\end{center}
\caption{(a-c) An illustration to show how to calculate the potential of mean force between two nanoparticles with the use of the pseudo-spring method: (a) two nanoparticles (big spheres) are linked by a spring (in $x$ axis) with binding ions (small spheres). (b) Fluctuation in separation $x$ between the centers of two nanoparticles with a spring in 0.1M 2:2 salt solution. The original length of the spring is 20\AA{}. (c) The distribution probability $p(x)$ (cyan line) at separation $x$ between the two nanoparticles from the statistical analysis on the data shown in (b). Solid line is fitted with the Gaussian function. (d-f) An illustration to show how to calculate the potential of mean force between two nanoparticles with the use of the inversed-Boltzmann method: (d) Fluctuation in separation $x$ between the centers of two nanoparticles \textit{versus} Monte Carlo steps. (e) The distribution probability $p(x)$ of the separation $x$ between the centers of nanoparticles from the 
statistical analysis on the data shown in (d). (f) The potential of mean force between the two nanoparticles
calculated from (e) with the inversed-Boltzmann method.
} \label{setup}
\end{figure}

\newpage
\begin{figure}[ht]
\begin{center}
\epsfxsize= 18.0cm \centerline{\epsfbox{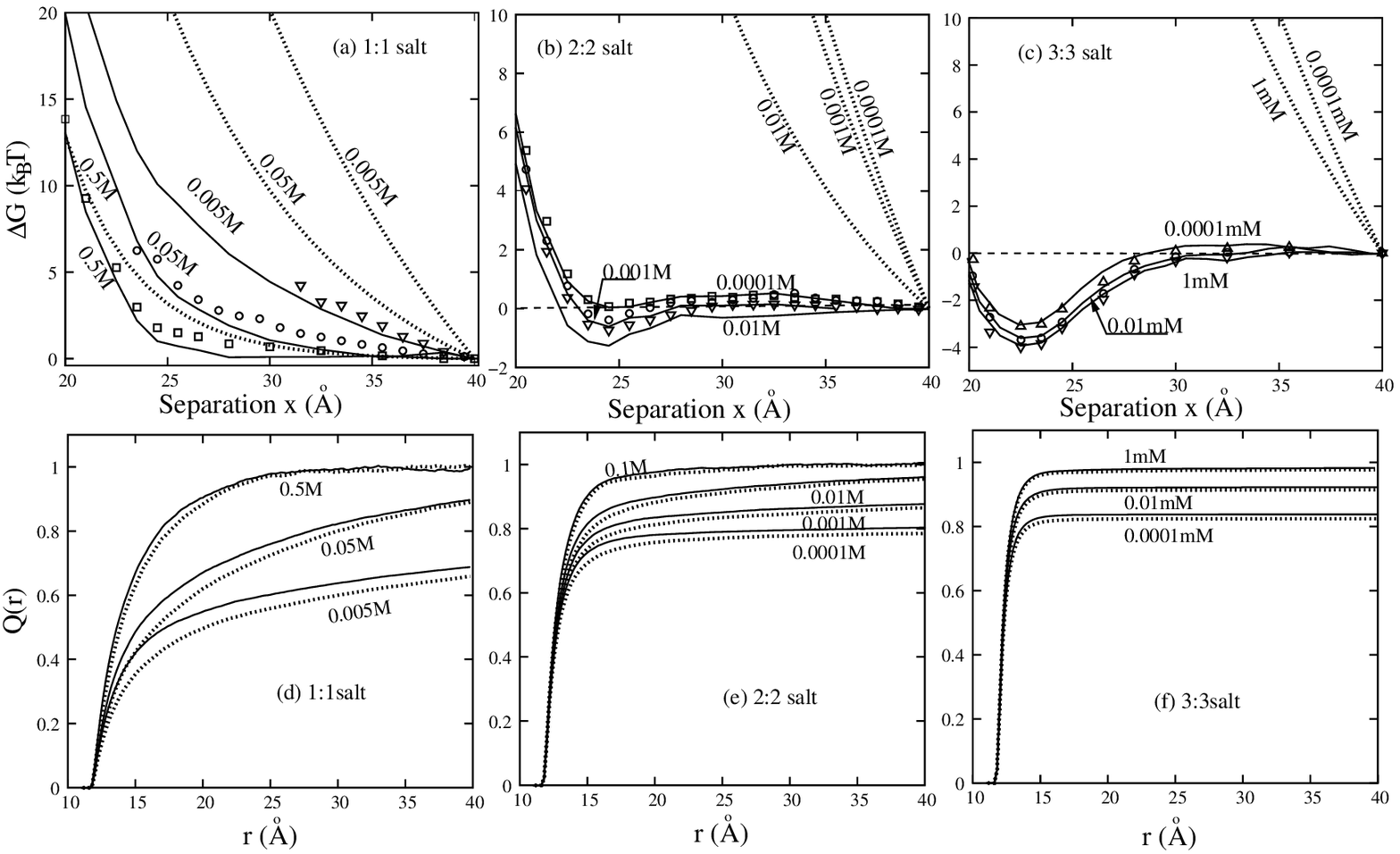}}
\end{center}
\caption{(a-c) The potentials of mean force as functions of the separation $x$ between two nanoparticles. (a) [1+]=0.5M ($ \square $), 0.05M ($ \bigcirc $) and 0.005M ($ \bigtriangledown $) (from the bottom to the top). (b) [2+]=0.01M ($\bigtriangledown$), 0.001M ($\bigcirc $) and 0.0001M ($\square$) (from the bottom to the top). (c) [3+]=1mM ($\bigtriangledown$), 0.01mM ($\bigcirc $) and 0.0001mM ($ \bigtriangleup $) (from the bottom to the top). Solid lines: calculated from the pseudo-spring method; Symbols: calculated from the inversed-Boltzmann method; Dotted lines: DLVO potentials (Eq. 1). (d-f) Net charge distribution $Q(r)$ per unit charge ($\rm{Eq.\,\, 8}$) on nanoparticles as a function of distance $r$ around the nanoparticles in 1:1 (d), 2:2 (e) and 3:3 (f) salt solutions. (d) [1+]=0.5M, 0.05M and 0.005M; (e) [2+]=0.01M, 0.001M and 0.0001M; (f) [3+]=1mM, 0.01mM and 0.0001mM.  Solid lines: the separation 
$x$'s between the centers of nanoparticles are 25\AA{} (d,e) and 23\AA{} (f) corresponding to the free energy minimum for 2:2 and 3:3 salt solution, respectively. Dashed lines: the separation $x$ is 40\AA{}, the outer
reference separation.
} \label{setup}
\end{figure}

\newpage
\begin{figure}[ht]
\begin{center}
\epsfxsize= 18cm \centerline{\epsfbox{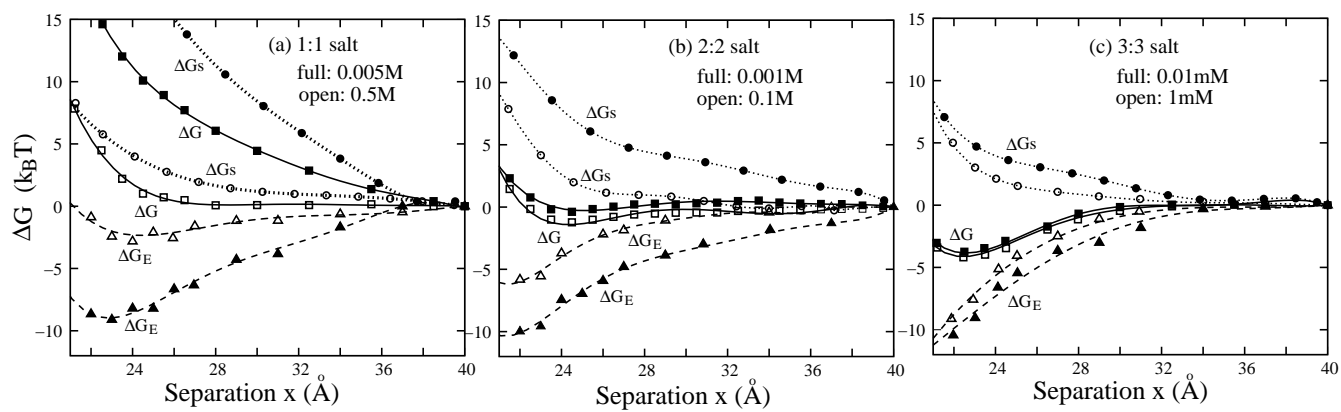}}
\end{center}
\caption{The potentials of mean force \(\rm \Delta G \) are composed of two contributions: electrostatic energy \(\rm \Delta G_E \) and entropic free energy \(\rm \Delta G_S \). (a) [1+]=0.5M and 0.005M; (b) [2+]=0.1M and 0.001M; (c) [3+]=0.001M and 0.00001M.
} \label{setup}
\end{figure}

\newpage
\begin{figure}[ht]
\begin{center}
\epsfxsize= 18cm \centerline{\epsfbox{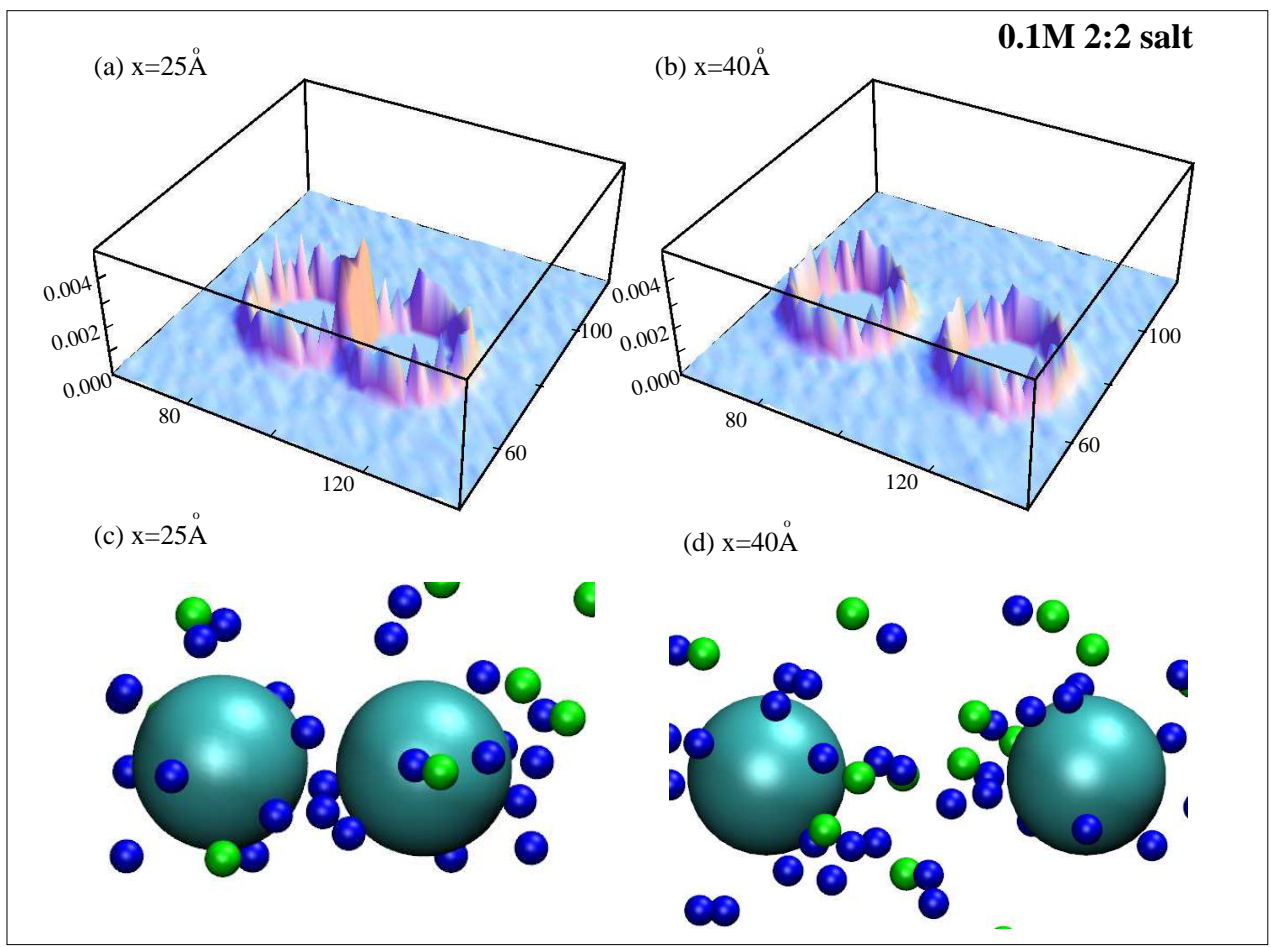}}
\end{center}
\caption{(a,b) The averaged net ion charge density (in unit of \( e/\textrm{\AA}{}^{3} \)) around two nanoparticles in  0.1M 2:2 salt solution at different separations: $x=25\textrm{\AA}{}$ (a) and $x=40\textrm{\AA}{}$ (b). (c,d) Snapshots show the structures of the cations (blue small spheres) and anions (green small spheres) around the two nanoparticles (big spheres) corresponding to system of (a) and (b), respectively.
} \label{setup}
\end{figure}

\newpage
\begin{figure}[ht]
\begin{center}
\epsfxsize= 18cm \centerline{\epsfbox{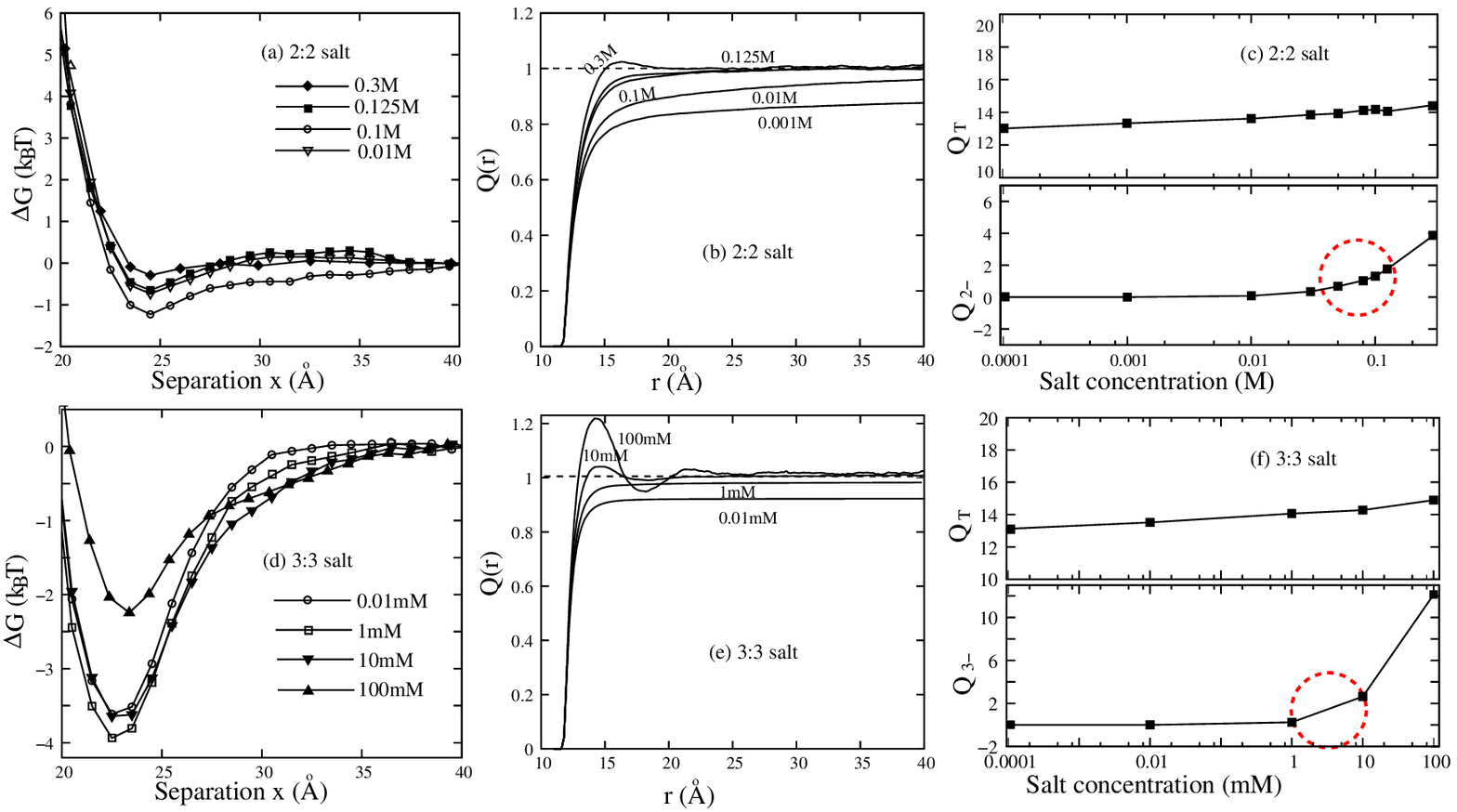}}
\end{center}
\caption{(a,d) The potential of mean force \(\rm \Delta G \) between two negatively charged nanoparticles as a function of the inter-nanoparticle separation $x$. (a) [2+]=0.01M, 0.1M, 0.125M and 0.3M; (d) [3+]=0.01mM, 1mM, 10mM and 100mM. The full-symbol lines are for the cases of (a) 0.125M and 0.3M [2+] and (d) 10mM and 100mM [3+]. (b,e) Net charge distribution $Q(r)$ per unit charge (Eq.\ref{Eq.9}) on nanoparticle around the two nanoparticles for the systems of (a) and (d) at $x=25{\textrm{\AA}}$. (c,f) The total ion charge distributions ($Q_{T}$) and negative ion charge distributions ($Q_{2-}$ and $Q_{3-}$) in the shaded region [see Fig. 6a] as functions of [2+] (c) and [3+] (f) at $x=25{\textrm{\AA}}$.
} \label{setup}
\end{figure}

\newpage
\begin{figure}[ht]
\begin{center}
\epsfxsize= 18cm \centerline{\epsfbox{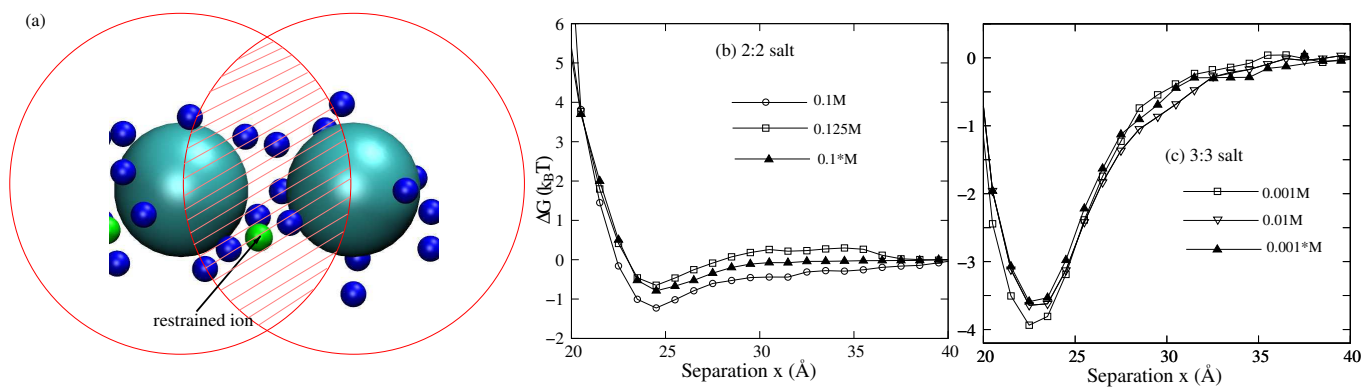}}
\end{center}
\caption{(a) An illustration to show how to restrain a anion in the inter-nanoparticle region denoted by the shaded region which is the overlapping area of the two circles, whose centers are those of the two nanoparticles, and radii both are $x$ (the separation between the two nanoparticles). (b,c) The potential of mean force between two negatively charged nanoparticles with a anion restrained in the shaded region in for 0.1M [2+] (b) and in 0.001M [3+] (c) solutions, which are denoted by $0.1^{\ast}$M and $0.001^{\ast}$M (full symbols), respectively. The lines with open symbols are the potentials of mean forces without a restrained anion in the shaded region.
} \label{setup}
\end{figure}

\newpage
\begin{figure}[ht]
\begin{center}
\epsfxsize= 18cm \centerline{\epsfbox{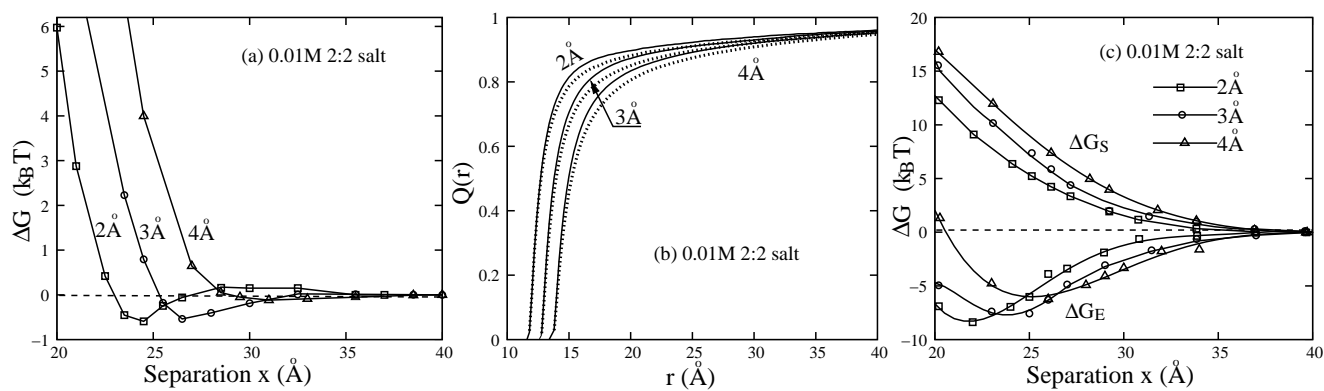}}
\end{center}
\caption{(a) The potential of mean force between two nanoparticles as a function of the inter-nanoparticle separation $x$ in 0.01M 2:2 salt  solution where the ion radii are 2\AA{}, 3\AA{}, and 4\AA{}, respectively. (b) Net charge distribution $Q(r)$ per unit charge on nanoparticle around the two nanoparticles for the systems of (a). Solid lines: $x=25\textrm{\AA}{}$. Dashed lines: $x=40\textrm{\AA}{}$.  (c) Electrostatic energy \(\rm \Delta G_E \) and entropic free energy \(\rm \Delta G_S \), corresponding to the potentials of mean force \(\rm \Delta G \) shown in (a).
} \label{setup}
\end{figure}

\newpage
\begin{figure}[ht]
\begin{center}
\epsfxsize= 18cm \centerline{\epsfbox{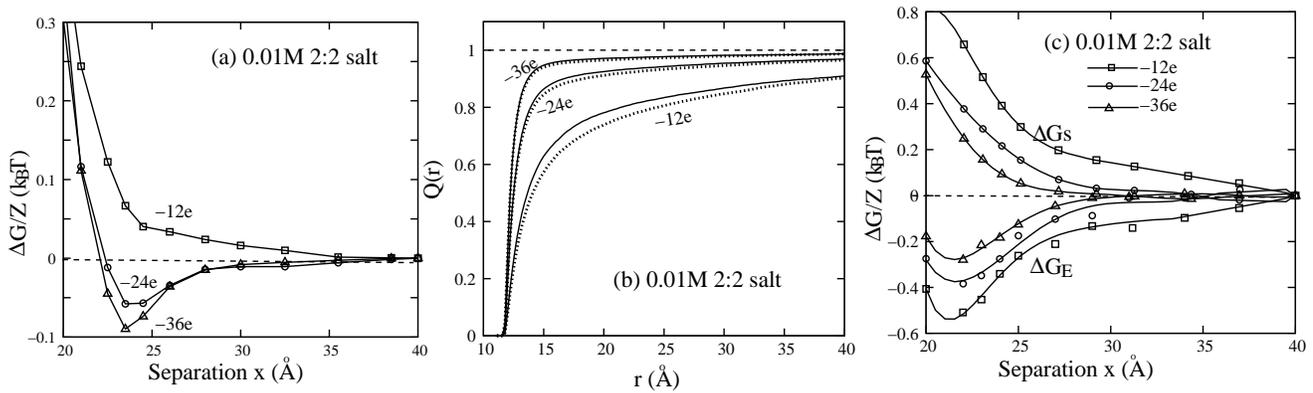}}
\end{center}
\caption{(a) The potential of mean force ($ \Delta$G/Z) per ($ -e $) between two nanoparticles as a function of the inter-nanoparticle separation $x$ in 0.01M 2:2 salt solution. The charges on each nanoparticle are $-36e$, $-24e$, and $-12e$, respectively. (b) Net charge distribution $Q(r)$ per unit charge on nanoparticle around two nanoparticles for the systems of (a). Solid lines: $x=25\textrm{\AA}{}$. Dashed lines: $x=40\textrm{\AA}{}$. (c) Electrostatic energy \(\rm \Delta G_{E}/Z \) and entropic free energy \(\rm \Delta G_{S}/Z \) per ($ -e $) correspond to the potentials of mean force \(\rm \Delta G \) shown in (a).
} \label{setup}
\end{figure}

\end{document}